\documentclass[conference]{IEEEtran}
\IEEEoverridecommandlockouts
\usepackage{cite}
\usepackage{amsmath,amssymb,amsfonts}
\usepackage{algorithmic}
\usepackage{graphicx}
\usepackage{booktabs}
\usepackage{textcomp}
\usepackage{xcolor}
\usepackage{url}
\usepackage{soul}
\def\BibTeX{{\rm B\kern-.05em{\sc i\kern-.025em b}\kern-.08em
    T\kern-.1667em\lower.7ex\hbox{E}\kern-.125emX}}
\begin{document}


\title{An opportunity to improve Data Center Efficiency: Optimizing the Server’s Upgrade Cycle\thanks{This work has been submitted and presented at the 1st International Workshop on Data Center Energy Efficiency (DCEE-2025) at ISCA-2025, June 21, 2025, Tokyo, Japan  }}

\author{\IEEEauthorblockN{Panagiota Nikolaou}
\IEEEauthorblockA{\textit{University of Central Lancashire, Cyprus}\\
}
\and
\IEEEauthorblockN{Freddy Gabbay }
\IEEEauthorblockA{\textit{The Hebrew University}\\
}
\and
\IEEEauthorblockN{Jawad Haj-Yahya}

\IEEEauthorblockA{\textit{Rivos Inc.}\\
}
\and
\IEEEauthorblockN{Yiannakis Sazeides}
\IEEEauthorblockA{\textit{University of Cyprus}\\
}

}

\maketitle
\vspace{-3cm}
\begin{abstract}
This work aims to improve a data center’s efficiency by optimizing the server upgrade plan: determine the optimal timing for replacing old servers with new ones. The opportunity presented by this approach is demonstrated through a study based on historical server data. The study establishes a significant opportunity to increase the $\frac{QPS}{(TCO \times CO2)}$ metric by formulating a global upgrade plan at the data center's design time covering its entire life cycle. This plan leverages information, such as server entry year, performance, and active power consumption
for both existing and future servers. Our findings reveal that an optimal global upgrade plan, may involve upgrades at non-fixed time periods and outperforms local upgrade plans. Local upgrade plans follow a fixed, equal-length cycle and make decisions based only on currently available server models. These local plans select the best available server at each upgrade cycle without accounting for future server releases.
\end{abstract}


\section{Introduction}
The modern hyperscale data centers have become significant contributors to the global energy consumption and CO\textsubscript{2} emissions
~\cite{IEA2022}. These trends are expected to continue and even become more prevalent in the future~\cite{freitag2021real}. At the same time, the performance and the total-cost-of-ownership (TCO)~\cite{hardy2013analytical} remain critical metrics in the design and operation of a data center, encapsulating the amount of work and the capital and operational expenses of a data center, respectively. Evidently, optimizing a modern data center is a multi-objective problem that requires trade-offs between competing metrics. 
Datacenter optimization has been focal to several studies~\cite{nikolaou2015modeling,kleanthous2015toward,jalili2021cost} and has been receiving renewed attention due to its large environmental footprint~\cite{wang2024designing,hanafy2024going,sudarshan2024eco,gupta2022act,patel2005cost}.

A data center policy, that has received little attention, is the server's upgrade cycle: how often to replace the servers in a data center~\cite{IntelServerRefresh, bashroush2018comprehensive}. Until recently, the server upgrade cycle was 3-4 years. Recently, major hyperscalar companies have announced increasing the upgrade cycles to 5-6 years as a way to address environmental concerns~\cite{miller2023meta,moss2023google}. 

In this work, we investigate the potential for improving a data center’s efficiency by optimizing its server upgrade cycle. In particular, we perform a study, based on historical server data, to establish the opportunity presented by the choice of the upgrade cycle to maximize the $\frac{QPS}{(TCO \times CO2)}$ metric. This is an exhaustive analysis of all possible upgrade plans.

For the servers considered in this study, the results reveal significant gains in the metric when a global upgrade plan— determines the optimal timing and selection of server replacements over data center's entire life cycle using both current and future server information—is derived at the design time of the data center. Such a plan is important to consider information (such as entry year, performance, active and idle power, CO\textsubscript{2}, cost, lifetime e.t.c.)  for both existing and future servers. Additionally, the findings show that for an optimal global upgrade plan, the updates occur with variable length time periods. Global plans are better than local plans; these are plans that i) have fixed equal-length upgrade cycles and ii) choose which new server to use, at each update cycle, based only on information for already available servers.

\begin{table*}[ht]
\centering
\caption{Server parameters\\ 
}
\label{tab:server_specs}
\resizebox{\textwidth}{!}{%
\begin{tabular}{lcccccccccccccccccccc}
\toprule
Server & Year & \#CPUs & \#DIMMS & CPU\_Cost & DIMM\_Price & QPS & Power & Util. & CI & Nr & Kr & Yield & Cifab & EPA & GPA & MPA & Area & CPSDram & CapDram \\
 & & & & (\$) & (\$) & & (W) & (\%) & (kgCO2/kWh) & & (kgCO2) & & (kgCO2/kWh) & (kWh/cm\textsuperscript{2}) & (kgCO2/cm\textsuperscript{2}) & (kgCO2/cm\textsuperscript{2}) & (cm\textsuperscript{2}) & (kgCO2/GB) & (GB) \\
\midrule
A & 2010 & 1 & 2 & 589 & 20 & 278441 & 130 & 100 & 0.23 & 1 & 0.15 & 0.5 & 0.295 & 0.41 & 0.14 & 0.5 & 2.96 & 0.6 & 4 \\
B & 2011 & 1 & 2 & 294 & 30 & 295443 & 58 & 100 & 0.23 & 1 & 0.15 & 0.5 & 0.295 & 0.793 & 0.17 & 0.5 & 2.16 & 0.6 & 8 \\
C & 2012 & 1 & 2 & 294 & 40 & 416999 & 59.5 & 100 & 0.23 & 1 & 0.15 & 0.5 & 0.295 & 1.08 & 0.18 & 0.5 & 1.6 & 0.315 & 8 \\
D & 2013 & 1 & 2 & 294 & 40 & 491887 & 58.5 & 100 & 0.23 & 1 & 0.15 & 0.5 & 0.295 & 1.08 & 0.18 & 0.5 & 1.6 & 0.315 & 8 \\
E & 2014 & 1 & 4 & 189 & 30 & 322278 & 131 & 100 & 0.23 & 1 & 0.15 & 0.5 & 0.295 & 1.08 & 0.18 & 0.5 & 1.77 & 0.315 & 16 \\
F & 2015 & 1 & 2 & 294 & 40 & 508794 & 60 & 100 & 0.23 & 1 & 0.15 & 0.5 & 0.295 & 1.08 & 0.18 & 0.5 & 1.6 & 0.315 & 16 \\
G & 2016 & 1 & 2 & 294 & 50 & 474667 & 47.9 & 100 & 0.23 & 1 & 0.15 & 0.5 & 0.295 & 1.2 & 0.2 & 0.5 & 1.22 & 0.065 & 16 \\
H & 2017 & 1 & 2 & 250 & 40 & 586973 & 56.1 & 100 & 0.23 & 1 & 0.15 & 0.5 & 0.295 & 1.2 & 0.2 & 0.5 & 1.22 & 0.065 & 16 \\
I & 2018 & 1 & 2 & 362 & 35 & 655699 & 71.7 & 100 & 0.23 & 1 & 0.15 & 0.5 & 0.295 & 1.2 & 0.2 & 0.5 & 1.54 & 0.065 & 16 \\
J & 2019 & 1 & 2 & 362 & 40 & 695687 & 61.8 & 100 & 0.23 & 1 & 0.15 & 0.5 & 0.295 & 1.2 & 0.2 & 0.5 & 1.54 & 0.065 & 16 \\
K & 2021 & 1 & 2 & 539 & 55 & 655851 & 55.5 & 100 & 0.23 & 1 & 0.15 & 0.5 & 0.295 & 1.2 & 0.2 & 0.5 & 2.76 & 0.065 & 16 \\
\bottomrule
\end{tabular}%
}

\footnotesize $Util$(Utilization),\footnotesize $N_r$(\# of ICs),\footnotesize $K_r$(IC packaging footprint),\footnotesize $Area$(IC Area),\footnotesize $MPA$(Procure materials),\footnotesize $EPA$(Fab energy),\footnotesize $CI$(HW CO\textsubscript{2} intensity),\footnotesize $CI_{fab}$(Fab CO\textsubscript{2} intensity),\footnotesize $GPA$(GHG from fab),\footnotesize $Yield$(Fab yield),\footnotesize $CPA$(Fab CO\textsubscript{2}),\footnotesize $CPSDram$(DRAM Carbon per size factor),\footnotesize $CapDram$(DRAM capacity)
\normalsize
\vspace{-0.5cm}
\end{table*}
%

\section{Methodology}
 \normalsize
A data center upgrade plan is defined by its start and end years. These correspond to the year a data center starts operation and the year that it is depreciated. A plan also includes an ordered vector with each element designating a year that servers get upgraded and the specific server used for the upgrade (for generality, we treat the first server acquisition as an upgrade). A server use period, $d$, is defined by the difference between the year a server is replaced (or reaches the data center depreciation period) and the year it got commissioned.

For each server in an upgrade plan, we first determine the TCO, performance, and CO$_2$ emissions for its use period, and then we aggregate each metric for all servers in the plan. 

To estimate the TCO of a server, we compute the sum of capital expenditures (CAPEX) and operational expenditures (OPEX). For the $CAPEX$ estimation, we consider the sum of the cost of all major server components (such as CPU and DIMMs).
For the $OPEX$ estimation, we use the following product: $Server\_Power * d * KiloHours/year * energy\_cost$.
%

For CO\textsubscript{2} estimation, we adopt the ACT model~\cite{gupta2022act}.
For the operational CO\textsubscript{2} emissions, we consider the server use period to account for the time that a server is active. 
It is important to note that if a depreciation plan involves reusing a specific server for multiple consecutive upgrade cycles, the CAPEX and embodied CO$_2$ emissions are incurred only once; no additional cost or emissions are added in subsequent cycles.

Finally, for the performance model, we assume each server is characterized by a throughput metric, like queries-per-second ($QPS$). The model determines the aggregate QPS by multiplying each server QPS by its use duration in seconds and then calculating the overall QPS by dividing all queries served over the data center's depreciation period by the depreciation period in seconds. 

This study is conducted using a dataset of 11 servers with one CPU that became available between 2010 and 2022 and reported their SPECpower score~\cite{specpower_ssj2008_results}. For the start year, we use 2010 and the end year 2022, representing a 12-year data center depreciation period. Each server's power consumption, performance (measured in operations per second), and number of DIMMs were derived from publicly available data, using the SPECpower benchmark under 100\% utilization~\cite{specpower_ssj2008_results}.

The cost input data were also obtained from publicly available sources, including major vendors such as Amazon and eBay, and CPU's chip area~\cite{intel_xeon_x3470}. For all other parameters used in the model, we adopted values from the ACT paper ~\cite{gupta2022act}, applying interpolation techniques when needed.
Table~\ref{tab:server_specs} shows all the input data for each of the 11 servers (A-K). 

For the exhaustive solver, we explore all possible upgrade plans to produce the best global plan. This analysis emulates a scenario that in the year 2010 of what someone could have decided if they had known the information for all 11 servers (i.e., even the future servers).

For the local upgrade plans, we assume decision-making is based only on the available servers up to each upgrade year (emulating the lack of future knowledge). We also limit local plans to fixed upgrade cycles of  1, 2, 3, 4, 6, or 12 years, thereby having the same upgrade cycle over the 12-year data center depreciation period.
\begin{figure}[t]
  \centering
  \includegraphics[width=1\linewidth]{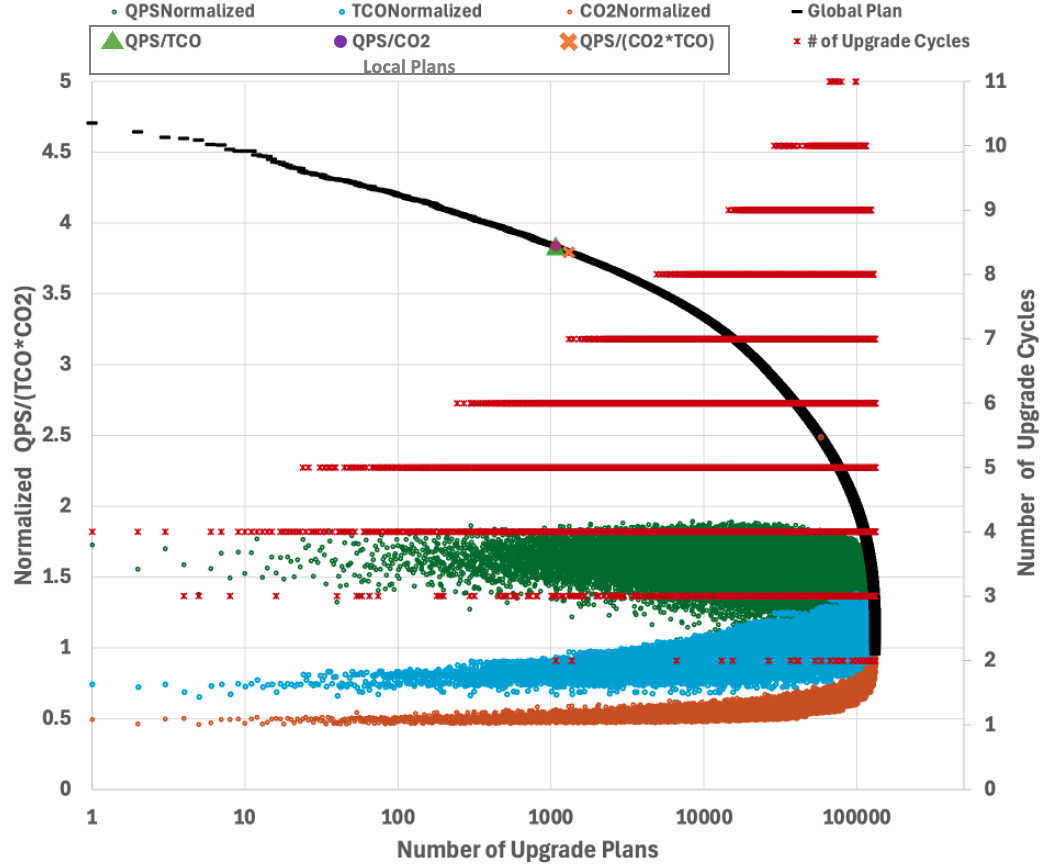}  
  \caption{Normalized Global and Local Plans for QPS/(TCO*CO2), TCO, CO2 and QPS}
  \vspace{-0.8cm}
  \label{fig:UpgradePlans}
\end{figure}

\section{Results}
Fig. \ref{fig:UpgradePlans} shows, for each global upgrade plan, on the x-axis, its QPS/(TCO × CO\textsubscript{2}) metric on the primary y-axis. The plans are sorted in descending order of the QPS/(TCO × CO\textsubscript{2}) metric.
The figure also points out which global plans correspond to the best local plans when optimizing them with the metrics QPS/TCO, QPS/CO$_2$, and QPS/(TCO × CO$_2$). Additionally, the primary y-axis shows the normalized QPS, TCO, and CO\textsubscript{2} values for each global plan. All values are normalized based on the plan that uses server $A$ throughout the entire 12-year data center lifetime, based on their respective metric. The secondary y-axis shows the number of upgrades for each plan.

As illustrated in Fig. \ref{fig:UpgradePlans}, the optimal global plan solution is 19\% better than the best local plan. This corresponds to the solution that uses the following 4 servers for the respective years: A: 1 year, B: 2 years, D: 4 years, H: 5 years. This solution does not offer neither the highest performance nor the lowest TCO or CO$_2$. These findings reveal a significant opportunity to improve the QPS/(TCO × CO$_2$) metric by deriving a global upgrade plan early in the lifetime of a datacenter. This encourages the use of information about future servers in the exploration and allows variable-length upgrade cycles. 
The graph, on the secondary y-axis, shows that the top ten solutions use 3 or 4 upgrade cycles. Moreover, these solutions converge to a small range of normalized TCO, CO$_2$, and performance. Such trends can be leveraged in a heuristic approach to navigate the upgrade plan design space faster. 


\vspace{-0.2cm}
\section{Conclusion and Future work}
This work shows that optimizing the server upgrade plan is a promising direction for increasing data center efficiency. A case study reveals that upgrade plans with variable cycle length that consider future server information can more effectively balance performance, cost, and environmental impact. This work motivates further investigation towards practical upgrade planning, such as:
1. predictive models for evolving hardware trends related to performance, power, reliability, and CO\textsubscript{2} emissions, that also handle the uncertainty of future servers.
2. more scalable, tractable, yet accurate solvers to address the complexity of exhaustive exploration, and
3. more realistic server lifetime models that account for hardware aging and degradation.
Another important direction is the implications of the optimization metric used.
\vspace{-0.2cm}
\section*{Acknowledgment}
This work is funded in part by the University of Cyprus research program RSHIELD. 
\vspace{-0.2cm}
\bibliographystyle{plain}
\bibliography{ref}

\begin{thebibliography}{10}

\bibitem{bashroush2018comprehensive}
Rabih Bashroush.
\newblock A comprehensive reasoning framework for hardware refresh in data centers.
\newblock {\em IEEE Transactions on Sustainable Computing}, 3(4):209--220, 2018.

\bibitem{freitag2021real}
Charlotte Freitag, Mike Berners-Lee, Kelly Widdicks, Bran Knowles, Gordon~S Blair, and Adrian Friday.
\newblock The real climate and transformative impact of ict: A critique of estimates, trends, and regulations.
\newblock {\em Patterns}, 2(9), 2021.

\bibitem{gupta2022act}
Udit Gupta, Mariam Elgamal, Gage Hills, Gu-Yeon Wei, Hsien-Hsin~S Lee, David Brooks, and Carole-Jean Wu.
\newblock Act: Designing sustainable computer systems with an architectural carbon modeling tool.
\newblock In {\em Proceedings of the 49th Annual International Symposium on Computer Architecture}, pages 784--799, 2022.

\bibitem{hanafy2024going}
Walid~A Hanafy, Qianlin Liang, Noman Bashir, Abel Souza, David Irwin, and Prashant Shenoy.
\newblock Going green for less green: Optimizing the cost of reducing cloud carbon emissions.
\newblock In {\em Proceedings of the 29th ACM International Conference on Architectural Support for Programming Languages and Operating Systems, Volume 3}, pages 479--496, 2024.

\bibitem{hardy2013analytical}
Damien Hardy, Marios Kleanthous, Isidoros Sideris, Ali~G Saidi, Emre Ozer, and Yiannakis Sazeides.
\newblock An analytical framework for estimating tco and exploring data center design space.
\newblock In {\em 2013 IEEE International Symposium on Performance Analysis of Systems and Software (ISPASS)}, pages 54--63. IEEE, 2013.

\bibitem{IntelServerRefresh}
{Intel Corporation}.
\newblock Data center efficiency: Realizing data center savings with accelerated server refresh.
\newblock \url{https://www.intel.com/content/dam/doc/performance-brief/data-center-efficiency-realizing-data-center-savings-with-accelerated-server-refresh-brief.pdf}, 2021.

\bibitem{IEA2022}
{International Energy Agency}.
\newblock Global trends in internet traffic, data centre workloads and data centre energy use, 2010--2019.
\newblock \url{https://www.iea.org/data-and-statistics/charts/global-trends-in-internet-traffic-data-centre-workloads-and-data-centre-energy-use-2010-2019}, 2022.

\bibitem{jalili2021cost}
Majid Jalili, Ioannis Manousakis, {\'I}{\~n}igo Goiri, Pulkit~A Misra, Ashish Raniwala, Husam Alissa, Bharath Ramakrishnan, Phillip Tuma, Christian Belady, Marcus Fontoura, et~al.
\newblock Cost-efficient overclocking in immersion-cooled datacenters.
\newblock In {\em 2021 ACM/IEEE 48th Annual International Symposium on Computer Architecture (ISCA)}, pages 623--636. IEEE, 2021.

\bibitem{kleanthous2015toward}
Marios Kleanthous, Yiannakis Sazeides, Emre {\"O}zer, Chrysostomos Nicopoulos, Panagiota Nikolaou, and Zacharias Hadjilambrou.
\newblock Toward multi-layer holistic evaluation of system designs.
\newblock {\em IEEE Computer Architecture Letters}, 15(1):58--61, 2015.

\bibitem{miller2023meta}
Rich Miller.
\newblock Meta will abandon some data center builds, run servers longer.
\newblock {\em Data Center Frontier}, February 2023.

\bibitem{moss2023google}
Sebastian Moss.
\newblock Google increases server life to six years, will save billions of dollars.
\newblock {\em Data Center Dynamics}, February 2023.

\bibitem{nikolaou2015modeling}
Panagiota Nikolaou, Yiannakis Sazeides, Lorena Ndreu, and Marios Kleanthous.
\newblock Modeling the implications of dram failures and protection techniques on datacenter tco.
\newblock In {\em Proceedings of the 48th International Symposium on Microarchitecture}, pages 572--584, 2015.

\bibitem{patel2005cost}
Chandrakant~D Patel and Amip~J Shah.
\newblock Cost model for planning, development and operation of a data center.
\newblock {\em Hewlett-Packard Laboratories Technical Report}, 107:1--36, 2005.

\bibitem{specpower_ssj2008_results}
{Standard Performance Evaluation Corporation}.
\newblock {SPECpower\_ssj2008 Results}.
\newblock \url{https://www.spec.org/power_ssj2008/results/}, 2025.

\bibitem{sudarshan2024eco}
Chetan~Choppali Sudarshan, Nikhil Matkar, Sarma Vrudhula, Sachin~S Sapatnekar, and Vidya~A Chhabria.
\newblock Eco-chip: Estimation of carbon footprint of chiplet-based architectures for sustainable vlsi.
\newblock In {\em 2024 IEEE International Symposium on High-Performance Computer Architecture (HPCA)}, pages 671--685. IEEE, 2024.

\bibitem{intel_xeon_x3470}
{TechPowerUp}.
\newblock Cpu spec database.
\newblock https://www.techpowerup.com/cpu-specs/, 2025.

\bibitem{wang2024designing}
Jaylen Wang, Daniel~S Berger, Fiodar Kazhamiaka, Celine Irvene, Chaojie Zhang, Esha Choukse, Kali Frost, Rodrigo Fonseca, Brijesh Warrier, Chetan Bansal, et~al.
\newblock Designing cloud servers for lower carbon.
\newblock In {\em 2024 ACM/IEEE 51st Annual International Symposium on Computer Architecture (ISCA)}, pages 452--470. IEEE, 2024.

\end{thebibliography}
\vspace{12pt}

\end{document}